# Economic thermodynamics

S.A. Rashkovskiy

*Ishlinsky Institute for Problems in Mechanics of the Russian Academy of Sciences, Vernadskogo Ave., 101/1 Moscow, 119526, Russia*

*E-mail: rash@ipmnet.ru, Tel. +7 9060318854*

**Abstract.** A thermodynamic approach to the description of economic systems and processes is developed. It is shown that there is a deep analogy between the parameters of thermodynamic and economic systems (markets); so each thermodynamic parameter can be associated with a certain economic parameter or indicator. The economic meaning of such primordially thermodynamic concepts as internal energy and temperature has been established. It is shown that many economic laws, which in economic theory are a generalization of the results of observations, or are based on the analysis of the psychology of the behavior of market actors, within the framework of economic thermodynamics can be obtained as the natural and formal results of the theory. In particular, we show that economic thermodynamics allows a natural description of such a phenomenon as inflation. The thermodynamic conditions of market equilibrium stability are derived and analyzed, as well as the Le Chatelier's principle as applied to economic systems.

## 1. Introduction

In recent decades, the understanding has come that the methods and theories developed in relation to physical systems can be successfully used to description of the systems outside physics, consisting of a large number of interacting elements. In such systems, specific mechanisms of interaction of elements become secondary, while certain collective properties come to the fore, which do not depend on the nature of the system, and should be the same for both physical and non-physical systems. It is these collective properties that determine the behavior of such systems as a whole.

An example of this is the application of the methods of thermodynamics and thermostatistics to the description of economic systems and processes, despite the fact that economics, as a theory, is fundamentally different from theories in physics, primarily in terms of its structure and principles of construction [1].

The methods of thermodynamics and thermostatistics as applied to the description of economic systems and processes were considered in papers [2-19]. So in works [2,3,9-11, 17, 19], analogs of the first and second laws of thermodynamics are introduced and analyzed in relation to economic systems. The economic analogue of the Carnot cycle is considered in works [2,3,7-11].

In works [2-17, 19], a new (for economics) quantitative indicator, temperature, is introduced and its economic meaning is discussed. The law of increasing entropy as applied to economic processes was apparently first discussed in [18]. In works [2-6,9,12,13,15-17], various economic systems and processes are analyzed by methods of statistical physics and the connection of economic temperature with probability distributions of wealth and income has been shown.

At the same time, it should be noted that in the cited works, thermodynamic methods and thermodynamic terminology were often used intuitively and formally, at the level of external analogies, without strict theoretical justification.

In work [19], we have shown that thermodynamics of markets can be constructed as a phenomenological theory, by analogy with how it is done in physics.

In particular, proceeding from general principles, we showed [19] that the market for a certain goods is described by the equation

$$\delta Q = dE + pdV - \mu dN \qquad (1)$$

which, both in form and in content, is analogous to the first law of thermodynamics, where $E$ is the amount of money available in the system (internal energy of system), $V$ is the amount of goods in the system (volume of the system), $p$ is mean price of goods in the system (pressure), $N$ is the number of elements (market actors), $\mu$ is the financial potential - a change in the amount of money in the system when the number of its elements changes per unit (due to migration, dissociation, recombination, etc.), $\delta Q$ is the heat - the amount of money that the elements of one system directly transfer to the elements of another system without buying and selling goods (for example, direct investments, dividend payments, cash gifts, donations, taxes, subsidies, loans, loan payments, etc.) and without changing the number of elements in the system.

The first law (1) shows that there are three ways to change the energy of an economic system [19]: (i) by changing the volume of the system (work); (ii) by changing the number of system elements (financial work); (iii) without changing the volume of the system and the number of its elements due to the direct transfer of money between the elements (heat transfer or heat exchange). There are no other ways to change the energy (the amount of money) of the economic system.

It was shown in [19] that for economic systems, the entropy $S$ can be introduced in a natural way, which for nonequilibrium processes satisfies the condition (second law)

$$dS \geq \delta Q/T \qquad (2)$$

where $T$ is the temperature of the economic system (economic temperature) - an intensive parameter that characterizes the economic system as a whole.

For equilibrium (quasi-static) processes, the entropy

$$S^0 = S^0(T, V, N) \qquad (3)$$

satisfies the equation (second law for equilibrium processes)

$$dS^0 = \delta Q/T \qquad (4)$$

As in physics, the parameters $(p, T, V, N, E)$ of the economic system are not independent, but are related by relations (equations of state) that characterize the given economic system [19]. These include the thermal equation of state of the economic system

$$p = f(T, V, N) \qquad (5)$$

and the energy (financial) equation of state of the economic system

$$E = E(T, V, N) \qquad (6)$$

which can be obtained by methods of economic thermostatistics [19].

In particular, for a primitive (ideal) market [19], the equation of state (5) has the form

$$pV = k_0 NT \qquad (7)$$

where $k_0$ is the numerical constant that determines the scale of temperature [19] – an analogue of the Boltzmann constant.

Despite the fact that economic temperature, as an indicator of the state of the economic system, was introduced and discussed in many works (see, for example, [2-17,19]), its real physical meaning and measurement methods remain unclear.

This paper is a further development of the ideas of the paper [19]. In particular, we will clarify the concepts of energy and temperature of an economic system, consider the thermodynamic conditions for the stability of the equilibrium of economic systems, and show the possibilities of thermodynamic methods in describing some economic processes.

**2. Energy of the economic system**

In [19], considering the thermodynamics of the market, we considered that the energy $E$ of the economic system is only the amount of money that is in this system. Let's consider this issue in more detail.

Suppose the amount of money that a person has is $M$. He decided to buy a car, the value of which is $U$. The amount of money that this person has left after buying a car is $M'$. Obviously, in the simplest case, we can write the equality ("the law of conservation of energy")

$$M = M' + U \qquad (8)$$

We assume here that there were no other expenses besides paying for the cost of the car when buying a car. If, when buying a car, there are any indirect costs (for example, a person had to pay some taxes, buy insurance, etc.), then they should be added as terms to the right hand side of equation (8).

After buying a car, a person's wealth will be the sum of the remaining money $M'$ and a car (for simplicity, we assume that a person does not own any other property or assets besides a car). Suppose that now this person wants to buy some commodity (for example, an apartment or a new car), the value of which is greater than the amount of money $M'$ he has. He, of course, can take out a bank loan. But let's say he doesn't want to take out a loan. Then he has a natural way out: to sell the car, add up his money $M'$ and the money received from the sale of the car, and buy the goods he is interested in. Thus, we can consider the car as a kind of potential source of money, which, however, cannot be directly used when buying a new goods; it must first be "converted" into money, and the money is already used to buy a new goods.

Consider another example. Let there are two people, which have the amount of money $M_1$ and $M_2$, respectively,. They decided to start a firm. To do this, they pooled their capitals and bought movable and immovable property (premises, equipment, etc.) necessary for the operation of the company. In addition, they deposited part of their money into the account of the company for its current activities. Let the amount of money in the account of the created company be equal to $M$, and the value of the movable and immovable property that the shareholders acquired for the company is equal to $U$. The amount of money that the shareholders have left after all these expenses are $M'_1$ and $M'_2$, respectively.

We can draw up a balance:

$$M_1 + M_2 = M'_1 + M'_2 + M + U \qquad (9)$$

As a result of such a recombination (combining simpler elements – people), a new element (firm) has appeared in the system (on the market), which in all economic processes acts as a single whole.

If we consider (9) as the "law of conservation of energy", we can say that the initial elements (people) have energy $\varepsilon_i = M_i$ ($i = 1,2$), while the company created by them has energy

$$\varepsilon = M + U \qquad (10)$$

We can also consider the reverse process when these shareholders decided to close the firm. In this case, they sell the firm (its business and its property), and divide the funds received among themselves. In this case, the law of conservation of energy has the form

$$M'_1 + M'_2 = M_1 + M_2 + M + U \qquad (11)$$

where $M_i$ and $M'_i$ are the amount of money the shareholders have before and after the sale of the company. This process can be considered as the dissociation (disintegration) of a complex element (firm) with the formation of two simpler elements (people).

Relations (8) - (11) show that monetary funds (cash and non-cash) play the role of kinetic energy in the economic system, while material resources (movable and immovable property, various assets, etc.) play the role of potential energy. Summarizing these considerations, we can say that potential energy in the economic system is something that cannot be used directly to paid when making purchases, but that, having sold, can be converted into money. Thus, the economic potential energy is converted into money only after the sale operation, which takes time and effort.

An analogue of example (8) in physics is a particle in a potential field. An analogue of example (9), (11) is a system of two particles interacting through potential forces, for example, a molecule consisting of two atoms. In the latter case, the "atoms" are two people united in a firm ("molecule"). The kinetic energies of atoms relative to the center of mass of the molecule are analogs of the funds $M_1$ and $M_2$ that remained with the shareholders after the creation of the company. The kinetic energy of a molecule associated with the movement of its center of mass is an analogue of the monetary funds of firm $M$, the potential energy of a molecule (the energy of interaction of atoms in a molecule, the binding energy of atoms in a molecule) is non-monetary material resources (assets) of firm $U$. When the molecule decays, stored in it potential energy is converted into kinetic energy of liberated atoms (in the case of exothermic decomposition) or, on the contrary, for the disintegration of a molecule, energy supply from the outside is required in order to break the bonds between atoms (in the case of endothermic decomposition of a molecule). Similarly, if a firm has a positive balance sheet (has no debts), then when it disintegrates, cash (kinetic energy) and money received from the sale of its material resources (potential energy) are transferred in the form of money (kinetic energy) to its shareholders. If the firm has a negative balance (for example, debts), then in order to close it (that is, "free the elements"), it is necessary to spend additional money (spend energy to break bonds, for example, supply energy from the outside).

Along with money, as well as movable or immovable property, there can be various financial instruments in the economic system, the sale or transfer of which ensures the receipt of funds. The difference between financial instruments and movable or immovable property is the depersonalization of financial instruments (like money), i.e. they are not registered, and when transferring them to another person, there is no need to re-register the ownership. At the same time, movable and immovable property is personified, i.e. when buying or selling it, you must re-register the ownership. For this reason, unlike movable or immovable property, financial instruments can be used as units of account, and, therefore, in certain conditions, they play the role of money.

Thus, the energy (wealth) of each element of the system consists of the amount of money, movable and immovable property, as well as the value of all financial instruments that this element has the ability to dispose of.

Generalizing these considerations to the entire economic system (market), we come to the conclusion that

$$E = M + \Gamma + U \quad (12)$$

plays the role of internal energy (wealth) of the system, where $M$ is the amount of money available to the elements of the system, $U$ is non-monetary material resources (movable and immovable property, etc.) available in the system, i.e. potential energy of the system, $\Gamma = \sum_s \Gamma^{(s)}$ is the amount of money in the system in the form of various financial instruments $\Gamma^{(s)}$, $s = 1, 2, \ldots \nu - 1$, $\nu$ is the number of financial instruments, available in the system, including money. It is obvious that financial instruments, together with money, play the role of the kinetic energy of the economic system

$$K = M + \Gamma \quad (13)$$

that is, energy that can be directly used to perform economic actions.

Then the total energy of system (12) can be written in the form

$$E = K + U \quad (14)$$

It should be borne in mind that in the general case $M \neq \sum_i M_i$, $\Gamma \neq \sum_i \Gamma_i$ and $U \neq \sum_i U_i$, where $M_i$, $\Gamma_i$ and $U_i$ are the amount of money, financial instruments and property (in monetary terms) that have different elements of the system; $i = 1, 2, \ldots N$, $N$ is the number of system elements. For example, for a system consisting of two elements (actors), in the general case, one can write $M = M_1 + M_2 - M_{12}$, where $M_1$ and $M_2$ are the amount of money that the first and second actor can dispose of, respectively, $M_{12}$ is the amount of money they can dispose of at the same time. The simple example in this case is the family. Similarly, for three elements one can write $M = M_1 + M_2 + M_3 - M_{12} - M_{13} - M_{23} + M_{123}$, where $M_{123}$ is the amount of money that all three elements can dispose of at the same time. In the general case, for $M$ one can use the well-known relation of set theory:

$$M \equiv M(A_1 \cup A_2 \cup \ldots \cup A_N) = M(A_1) + M(A_2) + \cdots + M(A_N) - M(A_1 \cap A_2) - \cdots - M(A_{N-1} \cap A_N) + \cdots + (-1)^{N-1} M(A_1 \cap A_2 \cap \ldots \cap A_N) \quad (15)$$

where $A_i$ means an element of the system. Likewise, one can write the potential energy of an economic system and the financial instruments it contains.

Taking into account the above, and repeating the reasoning of paper [19], we come to the conclusion that it is the total internal energy of system (12) that enters into the first law ("the law of conservation of energy") (1) and the canonical distribution.

## 3. Temperature of the economic system

Let us assume that the system has several financial instruments equivalent from an economic point of view, which the elements of the system can equally use in their economic activities. In this case, we can talk about different economic degrees of freedom. The number of economic degrees of freedom, including money, is equal to $\nu$. Using economic thermostatistics [19] and relation (12), it is easy to show that the mean energy of an equilibrium economic system with a constant number of elements (actors) of the system $N$ and a constant market volume $V$ is

$$E = N\nu k_0 T + \langle U \rangle \tag{16}$$

where $T$ is the equilibrium temperature of the economic system (economic temperature) – an intensive parameter in the canonical distribution that characterizes the equilibrium economic system as a whole; $\langle U \rangle$ is the mean potential energy of the market (the value of movable and immovable property of all actors in the economic system). Further, we will use the natural "energy" (economic) temperature scale, in which $k_0 = 1$ and the unit of measurement of the economic temperature is any monetary unit, for example, the US dollar, Euro, etc.

Calculation of $\langle U \rangle$ in (16) allows finding the energy (financial) equation of state (6) of an equilibrium economic system [19].

In particular, the energy of the primitive (ideal) market [19], for which $\langle U \rangle = 0, M = \sum_i M_i$, $\Gamma = \sum_i \Gamma_i$, is equal to

$$E = \nu NT \tag{17}$$

From the canonical distribution for the economic system [19], taking into account (12), it follows that

$$\langle M \rangle = \langle \Gamma^{(1)} \rangle = \langle \Gamma^{(2)} \rangle = \cdots = \langle \Gamma^{(\nu-1)} \rangle = NT \tag{18}$$

Relation (18) is a mathematical expression of the Equipartition theorem for an economic system: in an equilibrium (stable) economic system, money is uniformly distributed between all degrees of freedom of the economic system. It can be argued that the more uniform this distribution is in a real system, the more stable it is and the closer it is to equilibrium. This fact, obtained strictly within the framework of economic thermostatistics [19], is an expression of the famous saying: "Don't put all your eggs in one basket."

From relation (18), it follows that in the economic system, the role of temperature plays the ratio

$$T = \langle M \rangle / N \tag{19}$$

Thus, *the economic temperature is the mean amount of money per one element (actor) in the economic system.*

Economic temperature is as important an indicator of the state of an economic system as the temperature of a physical system is an essential indicator of the thermodynamic state of a gas, liquid or solid.

Economic temperature, as a new characteristic of the economic system, was considered in [2-17, 19], but its economic meaning remained unclear, the unit of measurement was unknown, and the measurement method was uncertain.

Expression (19) removes this uncertainty and indicates a natural method for measuring the temperature of an economic system. As a result, the energy (financial) equation of state (6) acquires physical meaning. In particular, if all the energy (wealth) of the economic system is concentrated in money (that is, there are no material values and other financial instruments besides money in the system), then $E = NT$, which corresponds to a primitive (ideal) market with one degree of freedom [19]. An analogue of such a primitive market in physical thermodynamics is an ideal gas, in which all particles have only one degree of freedom. For a primitive market with several degrees of freedom, relation (17) holds. In this case, the physical analogue of the primitive market is an ideal gas, the particles of which have $\nu$ degrees of freedom. If the market (gas) is not primitive (ideal), i.e. $\langle U \rangle \neq 0$, then its energy consists of both kinetic energy (money and financial instruments) and potential energy (movable and immovable property). In this case, the dependence of the energy of the economic system on temperature is no longer linear.

## 4. Simple thermodynamic processes in economic systems
### 4.1. Thermodynamic relations

The first law (1) describes an elementary thermodynamic process in an economic system, i.e. a process with a small change in the parameters of the system and a small thermal (in the economic sense [19]) impact. Taking into account the energy (financial) equation of state (6), the first law (1) can be rewritten as

$$\delta Q = \left(\frac{\partial E}{\partial T}\right)_{V,N} dT + \left[p + \left(\frac{\partial E}{\partial V}\right)_{T,N}\right] dV + \left[\left(\frac{\partial E}{\partial N}\right)_{T,V} - \mu\right] dN \quad (20)$$

Here, as is customary in thermodynamics, the subscript indicates the parameters that are considered constant when calculating the derivative.

Taking into account the thermal equation of state (5), we obtain

$$dV = \left(\frac{\partial V}{\partial p}\right)_{T,N} dp + \left(\frac{\partial V}{\partial T}\right)_{p,N} dT + \left(\frac{\partial V}{\partial N}\right)_{p,T} dN$$

Then equation (20) can be rewritten as

$$\delta Q = \left[\left(\frac{\partial E}{\partial T}\right)_{V,N} + \left(p + \left(\frac{\partial E}{\partial V}\right)_{T,N}\right)\left(\frac{\partial V}{\partial T}\right)_{p,N}\right]dT + \left[p + \left(\frac{\partial E}{\partial V}\right)_{T,N}\right]\left(\frac{\partial V}{\partial p}\right)_{T,N}dp + \left[\left(\frac{\partial E}{\partial N}\right)_{T,V} - \mu + \left(p + \left(\frac{\partial E}{\partial V}\right)_{T,N}\right)\left(\frac{\partial V}{\partial N}\right)_{p,T}\right]dN \qquad (21)$$

From equations (20) and (21), it follows that, in the general case, the thermodynamic process in an economic system is associated with heat exchange, with changes in temperature, volume, pressure and the number of elements in the system.

In what follows, we will consider the processes in an economic system with a constant number of elements ($dN = 0$).

In this case

$$\delta Q = \left(\frac{\partial E}{\partial T}\right)_{V,N} dT + \left[p + \left(\frac{\partial E}{\partial V}\right)_{T,N}\right]dV \qquad (22)$$

or

$$\delta Q = \left[\left(\frac{\partial E}{\partial T}\right)_{V,N} + \left(p + \left(\frac{\partial E}{\partial V}\right)_{T,N}\right)\left(\frac{\partial V}{\partial T}\right)_{p,N}\right]dT + \left[p + \left(\frac{\partial E}{\partial V}\right)_{T,N}\right]\left(\frac{\partial V}{\partial p}\right)_{T,N}dp \qquad (23)$$

For an economic system, as well as for physical systems, it is possible to introduce the concept of heat capacity: the amount of heat (the amount of money directly supplied to the system without the process of buying and selling goods) required to change the economic temperature of the system by one unit:

$$C = \frac{\delta Q}{dT} \qquad (24)$$

As follows from (24), in the general case, the heat capacity of an economic system is not a function of the state of the system, but is a characteristic of the economic process. In particular

$$C_V = \left(\frac{\delta Q}{dT}\right)_V, \quad C_p = \left(\frac{\delta Q}{dT}\right)_p \qquad (25)$$

is the heat capacity of the economic system at constant volume and constant pressure, respectively.

As in ordinary thermodynamics, in economic thermodynamics the second law in the form (1), (4) allows obtaining the general relations connecting different thermodynamic parameters of the system.

Taking into account (3) and (20), we can write the second law for equilibrium systems (4) in the form

$$\left(\frac{\partial S^0}{\partial T}\right)_{V,N} dT + \left(\frac{\partial S^0}{\partial V}\right)_{T,N} dV + \left(\frac{\partial S^0}{\partial N}\right)_{T,V} dN = \frac{\left(\frac{\partial E}{\partial T}\right)_{V,N} dT + \left[p + \left(\frac{\partial E}{\partial V}\right)_{T,N}\right]dV + \left[\left(\frac{\partial E}{\partial N}\right)_{T,V} - \mu\right]dN}{T} \qquad (26)$$

It follows from (26) that

$$\left(\frac{\partial S^0}{\partial T}\right)_{V,N} = \frac{1}{T}\left(\frac{\partial E}{\partial T}\right)_{V,N}, \quad \left(\frac{\partial S^0}{\partial V}\right)_{T,N} = \frac{1}{T}\left[p + \left(\frac{\partial E}{\partial V}\right)_{T,N}\right], \quad \left(\frac{\partial S^0}{\partial N}\right)_{T,V} = \frac{1}{T}\left[\left(\frac{\partial E}{\partial N}\right)_{T,V} - \mu\right] \qquad (27)$$

Applying the conditions $\frac{\partial^2 S^0}{\partial V \partial T} = \frac{\partial^2 S^0}{\partial T \partial V}$, $\frac{\partial^2 S^0}{\partial V \partial N} = \frac{\partial^2 S^0}{\partial N \partial V}$, $\frac{\partial^2 S^0}{\partial N \partial T} = \frac{\partial^2 S^0}{\partial T \partial N}$ to relations (27), after simple transformations we obtain the well-known thermodynamic relations [20]

$$T \left(\frac{\partial p}{\partial T}\right)_{V,N} = p + \left(\frac{\partial E}{\partial V}\right)_{T,N} \tag{28}$$

$$T \left(\frac{\partial \mu}{\partial T}\right)_{V,N} = \mu - \left(\frac{\partial E}{\partial N}\right)_{T,V} \tag{29}$$

$$\left(\frac{\partial p}{\partial N}\right)_{T,V} = -\left(\frac{\partial \mu}{\partial V}\right)_{T,N} \tag{30}$$

For a primitive (ideal) market, which is described by the equation of state (7), from condition (28) one obtains

$$\left(\frac{\partial E}{\partial V}\right)_{T,N} = 0 \tag{31}$$

That is, the energy of a primitive (ideal) market, which is at a constant economic temperature, does not depend on the amount of goods $V$ available on the market. This result is known in thermodynamics as Joule's second law [20].

In economic systems, as in physical systems, partial processes associated with the absence of some kind of impact are possible. We consider the main ones.

### 4.2. Isothermal process ($T = const$)

In economics, this is a process in which the mean amount of money per one element of the system does not change.

According to (22) and (23)

$$\delta Q = \left[p + \left(\frac{\partial E}{\partial V}\right)_{T,N}\right] dV \tag{32}$$

and

$$\delta Q = \left[p + \left(\frac{\partial E}{\partial V}\right)_{T,N}\right] \left(\frac{\partial V}{\partial p}\right)_{T,N} dp \tag{33}$$

This process is only possible as a result of heat exchange (for example, in the form of investment) and a simultaneous change in the volume or pressure of the system.

Because $\delta Q \neq 0$ for isothermal process, heat capacity $C = \infty$.

In economics, as in physics, one can introduce the concept of a thermostat or heat reservoir, i.e. a system with a very large supply of thermal energy (money). When the thermostat interacts with other systems, the thermodynamic state of the thermostat practically does not change, while the thermostat itself can independently change its thermodynamic state. Obviously, individual large banks can play the role of thermostats in the economy for a certain market segment. For economically independent states, their Central Banks are economic thermostats. For weak markets, the larger (strong) markets or other economically stronger states can be thermostats. So the United States is a kind of economic thermostat for the entire world, because it can supply "heat" to the common market through the emission of dollars, or can remove "heat", removing some of the dollars out of circulation.

### 4.3. A constant-pressure (isobaric) process ($p = const$)

In economic system, this is a process in which the price of a goods does not change.
Examples of an isobaric process are buying or selling goods at a constant price, while the quantity of the goods in the system (the volume of the system) changes. According to the first law (1), when a goods is sold ($dV < 0$), the useful work performed is used to increase the internal energy of the system and/or is spent on transferring heat to other systems (payments on loans, payments of dividends, taxes, etc.). When buying goods ($dV > 0$), the process can proceed either through the consumption of internal energy (own funds) of the system, or through heat received from the outside (through attracted funds: investments, loans, etc.).
According to (23) and (25), the heat capacity at constant pressure (in the isobaric process)

$$C_p = \left(\frac{\partial E}{\partial T}\right)_{V,N} + \left(p + \left(\frac{\partial E}{\partial V}\right)_{T,N}\right)\left(\frac{\partial V}{\partial T}\right)_{p,N} \tag{34}$$

In particular, for a primitive (ideal) market described by relations (7) and (17), one obtains

$$C_p = (\nu + 1)N \tag{35}$$

### 4.4. A constant-volume (isochoric) process ($V = const$)

In economics, this is a process in which the quantity of goods in the system does not change. In this case, useful work is not performed, and the internal energy (own funds) of the system changes only due to heat exchange with external systems, for example, by attracting funds from outside and accumulating them in the system, paying taxes, dividends, payments on loans, etc. without changing the amount of goods. In this case, according to (22) and (25), the heat capacity at constant volume is

$$C_V = \left(\frac{\partial E}{\partial T}\right)_{V,N} \tag{36}$$

From (34) and (36), it follows the relation connecting the heat capacities at constant volume and at constant pressure [20]:

$$C_p - C_V = \left(p + \left(\frac{\partial E}{\partial V}\right)_{T,N}\right)\left(\frac{\partial V}{\partial T}\right)_{p,N} \tag{37}$$

or, taking into account (28),

$$C_p - C_V = T \left(\frac{\partial p}{\partial T}\right)_{V,N} \left(\frac{\partial V}{\partial T}\right)_{p,N} \tag{38}$$

In particular, for a primitive (ideal) market described by relation (17), one obtains

$$C_V = \nu N \tag{39}$$

### 4.5. Polytropic process ($C = const$)

According to the definition (24), $\delta Q = C dT$.

Using (22), one obtains

$$C dT = \left(\frac{\partial E}{\partial T}\right)_{V,N} dT + \left[p + \left(\frac{\partial E}{\partial V}\right)_{T,N}\right] dV \tag{40}$$

Taking into account (36) and (37), we obtain

$$(C - C_V) dT = \frac{(C_p - C_V)}{\left(\frac{\partial V}{\partial T}\right)_{p,N}} dV \tag{41}$$

Because $C \neq C_V$ ($V \neq const$), then

$$dT = \frac{(C_p - C_V)}{(C - C_V)} \left(\frac{\partial T}{\partial V}\right)_{p,N} dV \tag{42}$$

This equation is a differential equation of a polytropic process in an economic system in variables $T$ and $V$.

From the equation of state of system (5), one can find the dependence $T = T(p, V, N)$. For $N = const$, one obtains

$$dT = \left(\frac{\partial T}{\partial V}\right)_{p,N} dV + \left(\frac{\partial T}{\partial p}\right)_{V,N} dp$$

Substituting this expression in (42), we obtain the differential equation of the polytropic process in the economic system in the variables $p$ and $V$:

$$\frac{dp}{p} + n \frac{dV}{V} = 0 \tag{43}$$

where

$$n = \frac{(C_p - C)}{(C_V - C)} \frac{V \left(\frac{\partial T}{\partial V}\right)_{p,N}}{p \left(\frac{\partial T}{\partial p}\right)_{V,N}} \tag{44}$$

is the polytropic index of the economic process.

Using the equation of state (5), we obtain

$$\left(\frac{\partial p}{\partial V}\right)_{T,N} \left(\frac{\partial V}{\partial T}\right)_{p,N} \left(\frac{\partial T}{\partial p}\right)_{V,N} = -1$$

Then relation (44) takes the form

$$n = -\frac{(C_p - C)}{(C_V - C)} \frac{V}{p} \left(\frac{\partial p}{\partial V}\right)_{T,N} \tag{45}$$

With a constant polytropic index, from equation (43) one obtains

$$pV^n = const \tag{46}$$

In the particular case of a primitive (ideal) market [19], described by the equation of state (7), one obtains

$$n = \frac{(C_p - C)}{(C_V - C)} \tag{47}$$

In economic analysis, parameters called elasticity are widely used, which show the sensitivity of some economic indicators to changes in others [1]. One of these indicators is Price elasticity measures the responsiveness of the quantity demanded or supplied of a good to a change in its price:

$$\beta = \frac{p}{V}\frac{dV}{dp} \tag{48}$$

From equation (43), it follows that for a polytropic economic process

$$\beta = -n^{-1} \tag{49}$$

### 4.6. Adiabatic process ($\delta Q = 0$)

In this process, the economic system does not exchange heat with other systems, i.e. does not attract money from outside in the form of investments, loans, etc. and does not give money to other systems in the form of dividends, taxes, loan payments, etc. So an adiabatic economic system can change its internal energy (own funds) in only one way: by buying or selling goods; there are no other ways of exchanging energy (money) with other (external) systems for an adiabatic system.

The adiabatic process is a special case of the polytropic process at $C = 0$, therefore, all the results obtained above for the polytropic process also apply to the adiabatic one.

From (45), for the adiabatic process one obtains

$$n = -k\frac{V}{p}\left(\frac{\partial p}{\partial V}\right)_{T,N} \tag{50}$$

where

$$k = C_p/C_V \tag{51}$$

is the adiabatic exponent.

In particular, for a primitive (ideal) market [19]

$$n = k \tag{52}$$

In this case, taking into account (35) and (39), one obtains

$$k = \frac{\nu + 1}{\nu} \tag{53}$$

Taking into account that the number of economic degrees of freedom $\nu \geq 1$, it follows from relation (53) that for a primitive (ideal) $1 < k \leq 2$.

## 5. Thermodynamic potentials of the economic system

As in conventional thermodynamics, various thermodynamic potentials can be introduced in economic thermodynamics, which can be useful in the analysis of economic systems.

For nonequilibrium processes in economic systems, the second law has the form (2).

Taking into account that the economic temperature is always positive by definition, using equation (1), we rewrite inequality (2) as

$$TdS \geq dE + pdV - \mu dN \tag{54}$$

Introducing the Helmholtz free energy of the economic system

$$F = E - TS \tag{55}$$

we rewrite inequality (54) as

$$dF \leq -pdV - SdT + \mu dN \tag{56}$$

As follows from (56), in the equilibrium state of the economic system, its Helmholtz free energy is a function of the parameters $T, V$ and $N$:

$$F^0 = F^0(T, V, N) \tag{57}$$

In this case, according to (56)

$$p = -\left(\frac{\partial F^0}{\partial V}\right)_{T,N}, \quad S^0 = -\left(\frac{\partial F^0}{\partial T}\right)_{V,N}, \quad \mu = \left(\frac{\partial F^0}{\partial N}\right)_{T,V} \tag{58}$$

As in conventional thermostatistics, the Helmholtz free energy (57) of an economic system (market) can be calculated using the partition function [19]. This provides a universal algorithm for constructing the thermodynamics of markets, and opens the way for the development of thermodynamic models of various economic systems.

Introducing the enthalpy of an economic system

$$H = E + pV \tag{59}$$

we rewrite inequality (54) as

$$dH \leq TdS + Vdp + \mu dN \tag{60}$$

Obviously, the enthalpy of an economic system is the wealth of the economic system plus the value of all the goods in the system.

As follows from (60), in the equilibrium state of the economic system, its enthalpy is a function of the parameters $S$, $p$ and $N$:

$$H^0 = H^0(S^0, p, N) \tag{61}$$

In this case, according to (60)

$$T = \left(\frac{\partial H^0}{\partial S^0}\right)_{p,N}, \quad V = \left(\frac{\partial H^0}{\partial p}\right)_{S^0,N}, \quad \mu = \left(\frac{\partial H^0}{\partial N}\right)_{S^0,p} \tag{62}$$

Introducing Gibbs free energy of the economic system

$$G = E + pV - TS \tag{63}$$

we rewrite inequality (54) as

$$dG \leq Vdp - SdT + \mu dN \tag{64}$$

As follows from (64), in the equilibrium state of the economic system, its Gibbs free energy is a function of the parameters $T, p$ and $N$:

$$G^0 = G^0(T, p, N) \tag{65}$$

In this case, according to (64)

$$V = \left(\frac{\partial G^0}{\partial p}\right)_{T,N}, \quad S^0 = -\left(\frac{\partial G^0}{\partial T}\right)_{p,N}, \quad \mu = \left(\frac{\partial G^0}{\partial N}\right)_{T,p} \tag{66}$$

Introducing the Grand Potential of the economic system

$$\Omega = E - TS - \mu N \tag{67}$$

we rewrite inequality (54) as

$$d\Omega \leq -pdV - SdT - Nd\mu \tag{68}$$

As follows from (68), in the equilibrium state of the economic system, its Grand potential is a function of the parameters $T, V$ and $\mu$:

$$\Omega^0 = \Omega^0(T, V, \mu) \tag{69}$$

In this case, according to (68)

$$p = -\left(\frac{\partial \Omega^0}{\partial V}\right)_{T,\mu}, \quad S^0 = -\left(\frac{\partial \Omega^0}{\partial T}\right)_{V,\mu}, \quad N = -\left(\frac{\partial \Omega^0}{\partial \mu}\right)_{T,V} \tag{70}$$

By definition, all thermodynamic potentials of an economic system are extensive parameters. This allows writing the relations (57), (61), (65), and (69) in the form

$$F^0 = Nf^0(T, V/N) \tag{71}$$
$$H^0 = Nh^0(S^0/N, p) \tag{72}$$
$$G^0 = Ng^0(T, p) \tag{73}$$
$$\Omega^0 = V\omega^0(T, \mu) \tag{74}$$

where $f^0, h^0, g^0$ and $\omega^0$ are the functions depending on only two intensive arguments.

Using the last relation (66) and relation (73), one obtains $g^0(T, p) = \mu(T, p)$. Then the Gibbs free energy (73) of an economic system in an equilibrium state can be written in the form

$$G^0 = N\mu(T, p) \tag{75}$$

Similarly, using the first relation (70) and relation (73), one obtains $\omega^0(T, \mu) = -p(T, \mu)$. Then the Grand potential (74) of an economic system in equilibrium can be written as

$$\Omega^0 = -Vp(T, \mu) \tag{76}$$

It can be expected that, as in ordinary thermodynamics, in economic thermodynamics, thermodynamic potentials will play an important role in the analysis of the equilibrium states of the economic system and the processes occurring in it. In particular, we see that if the dependence (57) of the Helmholtz free energy is known, then the equilibrium price of the goods is found by simple differentiation of $F^0$ with respect to $V$, and if the Grand potential of the economic system is known, then the equilibrium price of the goods can be found directly from the relation (76). Similarly, if the dependence (65) of the Gibbs free energy is known, then the equilibrium quantity of goods in the market corresponding to a given mean price $p$ can be found by differentiating $G^0$ with respect to $p$. In particular, taking into account (75), we rewrite the first relation (66) in the form

$$V = N \left(\frac{\partial \mu}{\partial p}\right)_T \tag{77}$$

Relation (77), in fact, is another form of writing the thermal equation of state (5) of the economic system.

On the other hand, knowing the thermal equation of state (5) of the economic system (market), using equation (77), one can find the financial potential of the market $\mu(T,p)$ up to an arbitrary function of temperature. For example, for a primitive (ideal) market described by the equation of state (7), one writes (77) in the form

$$\left(\frac{\partial \mu}{\partial p}\right)_T = T/p \tag{78}$$

The solution to equation (78) has the form

$$\mu(T,p) = T \ln(p/p_\mu) \tag{79}$$

where $p_\mu = p_\mu(T)$ is the arbitrary function of temperature.

Taking into account the obvious relationship

$$\left(\frac{\partial \mu}{\partial T}\right)_{V,N} = \left(\frac{\partial \mu}{\partial T}\right)_p + \left(\frac{\partial \mu}{\partial p}\right)_T \left(\frac{\partial p}{\partial T}\right)_{V,N}$$

one writes relation (29) in the form

$$T \left(\frac{\partial \mu}{\partial T}\right)_p + T \left(\frac{\partial \mu}{\partial p}\right)_T \left(\frac{\partial p}{\partial T}\right)_{V,N} = \mu - \left(\frac{\partial E}{\partial N}\right)_{T,V} \tag{80}$$

Substituting relations (7), (17), and (79) into equation (80), one obtains

$$\frac{T}{p_\mu} \frac{dp_\mu}{dT} = (\nu + 1) \tag{81}$$

The solution to equation (81) has the form

$$p_\mu = p_1 (T/T_1)^{\nu+1} \tag{82}$$

where $p_1$ and $T_1$ are the arbitrary constants.

Thus, the financial potential of the primitive (ideal) market [19] is determined by relations (79) and (82) up to an arbitrary constant $p_1/T_1^{\nu+1}$.

Using the equation of state (7), one writes the financial potential (79), (82) in the form

$$\mu(T,p) = T(\ln x - \nu \ln T + B) \tag{83}$$

where $B = \ln(T_1^{\nu+1}/p_1)$ is the constatnt;

$$x = N/V \tag{84}$$

Is the "concentration" of elements in the system (the number of elements per unit of goods).

## 6. Thermodynamic theory of inflation

In economics, inflation is a general rise in the price level in an economy over a period of time [1].

We can talk about inflation (rise in prices) for specific goods, for a specific group of goods, or for the economy as a whole.

Inflation for an individual goods is characterized by the parameter $\pi = \dot{p}/p$ – the relative rate of change in the price of the goods.

The existing theories of inflation [21-24] and others are phenomenological in nature, and are based on more or less well-grounded assumptions about the role of various factors in the process of rising prices.

As is known [22], one of the components of inflation is the expected inflation $\pi^e$, which sellers are guided by when changing the price of goods.

The change in expected inflation is described by the relaxation-type equation underlying the well-known Kagan model [22]:

$$\dot{\pi}^e = \gamma(\pi - \pi^e) \tag{85}$$

where $\gamma > 0$ is the adaptive inflation expectations parameter.

People (buyers and sellers), when choosing a strategy of behavior, are guided by their forecasts regarding expected inflation. This is described by equation (85), which takes into account the lag of expectations compared to reality. The lag between reality and expectation will be the greater, the faster real inflation changes, i.e., the more $\dot{\pi}$. With a slow (quasi-static) change in real inflation ($\dot{\pi}/\gamma \ll \pi$), the expected inflation will be approximately equal to the real one: $\pi^e \approx \pi$. In this case, we can talk about equilibrium (quasi-static) inflation.

Thus, two components of inflation can be distinguished: (i) equilibrium (thermodynamic inflation) and (ii) nonequilibrium.

Equilibrium inflation corresponds to a slow (quasi-static) change in inflation and, accordingly, inflation expectations, in time when the expected inflation is approximately equal to real inflation. The nonequilibrium component manifests itself with a rapid change in inflation over time, when the society (system) does not have time to adapt to the existing inflation at every moment of time.

Economic thermodynamics allows explaining and describing the quasi-static (equilibrium) inflation.

Consider the simple case when the quantity of goods and the number of market actors do not change ($V = const, N = const$), i.e. an isochoric economic process takes place.

In this case, from the equation of state of the market (5), one obtains

$$\frac{dp}{p} = \left(\frac{\partial \ln p}{\partial \ln T}\right)_{V,N} \frac{dT}{T} \qquad (86)$$

Taking into account the definition of economic temperature (19), which for a specific system can be rewritten as

$$T = M/N \qquad (87)$$

equation (86) can be written in the form of the simple inflation equation

$$\pi = am \qquad (88)$$

where $m = \dot{M}/M$ is the relative rate of change in the amount of money in the system;

$$a = \left(\frac{\partial \ln p}{\partial \ln T}\right)_{V,N} \qquad (89)$$

For a primitive (ideal) market [19], the equation of state of which has the form (7), in the case under consideration one obtains $a = 1$ and $\pi = m$. Thus, in this case, inflation is associated only with the growth of the money supply in the system, and the rate of inflation is equal to the rate of growth of the money supply.

In more general case, inflation can occur with variations in $V$ and $N$, when the quantity of goods and the number of participants in the system change. In this case, to describe inflation, it is necessary to consider the first law (1) taking into account the real process and the equations of state. In particular, if the process occurring in the economic system is polytropic at $N = const$, then, taking into account (42) - (44), we again obtain the inflation equation (88) with

$$a = \frac{(C_p - C)}{(C_p - C_V)} \left(\frac{\partial \ln p}{\partial \ln T}\right)_{V,N} \qquad (90)$$

For a primitive (ideal) market [19], in this case, one obtains

$$a = \frac{C_p - C}{C_p - C_V} \qquad (91)$$

or

$$a = \frac{n}{n-1} \qquad (92)$$

where $n$ is the polytropic index (47) of the economic process in the system.

In particular, inflation can occur even in an adiabatically isolated (from an economic point of view [19]) economic system, i.e. without pumping it with money, not related to the sale and purchase of goods: $\delta Q = 0$. In this case, inflation is associated with the fact that the quantity of goods in system $V$ increases more slowly than the amount of money in the system, or, conversely, the quantity of goods in the system decreases faster than the amount of money.

In the general case, equilibrium (quasi-static) inflation is described by the first law (1) in its various forms.

Using the first law in the form (21), one obtains

$$\pi = a\tau - b_\tau \omega - \sigma q \tag{93}$$

where

$$\tau = \dot{T}/T \tag{94}$$

$$\omega = \dot{N}/N \tag{95}$$

$$q = \frac{\dot{Q}}{NT} \tag{96}$$

$$a = -\frac{T\left(\frac{\partial E}{\partial T}\right)_{V,N} + \left(p + \left(\frac{\partial E}{\partial V}\right)_{T,N}\right)T\left(\frac{\partial V}{\partial T}\right)_{p,N}}{\left[p + \left(\frac{\partial E}{\partial V}\right)_{T,N}\right]p\left(\frac{\partial V}{\partial p}\right)_{T,N}} \tag{97}$$

$$b_\tau = \frac{N\left(\frac{\partial E}{\partial N}\right)_{T,V} - \mu N + \left(p + \left(\frac{\partial E}{\partial V}\right)_{T,N}\right)N\left(\frac{\partial V}{\partial N}\right)_{p,T}}{\left[p + \left(\frac{\partial E}{\partial V}\right)_{T,N}\right]p\left(\frac{\partial V}{\partial p}\right)_{T,N}} \tag{98}$$

$$\sigma = -\frac{NT}{\left[p + \left(\frac{\partial E}{\partial V}\right)_{T,N}\right]p\left(\frac{\partial V}{\partial p}\right)_{T,N}} \tag{99}$$

The inflation equation (93) establishes a relationship between the rate of price growth $\pi$ and the rate of change in economic temperature (94) at equilibrium (quasi-static) inflation.

Usually, when analyzing inflation, one analyzes the relationship between the inflation rate (the rate of price growth $\pi$) and the rate of growth of the money supply in the system $m = \dot{M}/M$, which, taking into account (87), can be written in the form

$$\dot{M}/M = \dot{T}/T + \dot{N}/N \tag{100}$$

Using equation (93) and relation (100), one writes the equation for equilibrium (quasi-static) inflation in the form

$$\pi = am - b\omega - \sigma q \tag{101}$$

where

$$b = \frac{N\left(\frac{\partial E}{\partial N}\right)_{T,V} - T\left(\frac{\partial E}{\partial T}\right)_{V,N} - \mu N + \left(p + \left(\frac{\partial E}{\partial V}\right)_{T,N}\right)\left(N\left(\frac{\partial V}{\partial N}\right)_{p,T} - T\left(\frac{\partial V}{\partial T}\right)_{p,N}\right)}{\left[p + \left(\frac{\partial E}{\partial V}\right)_{T,N}\right]p\left(\frac{\partial V}{\partial p}\right)_{T,N}} \tag{102}$$

In particular, for a primitive (ideal) market [19], one obtains

$$a = 2, b_\tau = \frac{\mu}{T} - 2, b = \frac{\mu}{T}, \sigma = 1 \tag{103}$$

or, taking into account (79),

$$b_\tau = \ln(p/p_\mu) - 2, b = \ln(p/p_\mu) \tag{104}$$

Then equations (93) and (101) take the form

$$\pi = 2\tau - \left(\ln(p/p_\mu) - 2\right)\omega - q \tag{105}$$

and

$$\pi = 2m - \omega \ln(p/p_\mu) - q \tag{106}$$

Taking into account that $a > 0$, it follows from (101) that an increase in the amount of money in the system ($m > 0$), other things being equal, leads to an increase in prices (i.e., to inflation). Taking into account that $\sigma > 0$, it follows from (93) and (101) that investments have a positive effect on the inflation rate: the greater the inflow of investments (($\dot{Q} > 0$) into the system, the lower the inflation rate. On the contrary, if money is withdrawn from the system (($\dot{Q} < 0$), for example, the tax burden increases, investments are withdrawn, etc., then inflation accelerates, and the more $|\dot{Q}|$, the stronger. These conclusions are fully consistent with existing ideas about inflation [1].

Consider the role of competition in the market in inflation process. Taking into account that $N$ is the number of market actors, we come to the conclusion that with an increase in $N$, competition in the market grows, and the parameter $\omega$ (95) describes the rate of growth of competition in the market. It follows from equation (101) that an increase in the number of market actors (i.e., increase in competition in the market; $\omega > 0$) with $b > 0$ and other conditions being equal (fixed $m$ and $q$) leads to a decrease in inflation, while with $b < 0$ – to its grows. For $b > 0$, this conclusion is quite obvious from a psychological point of view: all sellers are fighting for a buyer and strive to reduce the price of the goods in order to attract more buyers. From equations (104) and (106) it follows that for a primitive (ideal) market with goods price $p > p_\mu$, the parameter $b > 0$, while with goods price $p < p_\mu$, the parameter $b < 0$. So the positive role of an increase in the number of market actors (competition) in inflation process is manifested only for $p > p_\mu$, but it is negative for $p < p_\mu$, when increased competition in the market leads to an increase in inflation. This result is unusual and requires further analysis. For example, we can assume that for markets always, $p > p_\mu$, i.e. the price of goods cannot be lower than a certain threshold value $p_\mu$ (e.g., the price of goods cannot be lower than its cost price).

Before that, we considered the change in the price of a particular goods – "partial" inflation. However, in practice, inflation is estimated not for each individual product, but, for example, for a market with several types of goods or even for the economy as a whole [1]. For this purpose, a market basket (commodity bundle) is formed – a fixed list of items, in given proportions [1].

The market basket, which is used to estimate the rate of inflation, can be compared to a thermodynamic system consisting of parallel working cylinders with pistons (see Fig. 1). Each cylinder with a piston is a separate type of goods included in the market basket. The gas filling each of the cylinders under pressure $p_i$ is the market for the corresponding goods with the mean price $p_i$, $i = 1, \ldots r$, $r$ is the number of cylinders in the system (the number of goods in the commodity bundle). The force acting from all pistons on the "market basket" plate is the price of the commodity bundle $R = \sum_{i=1}^{r} p_i S_i$, where $S_i$ is the area of the $i$-th piston – the number of the $i$-th of the goods included in the market basket.

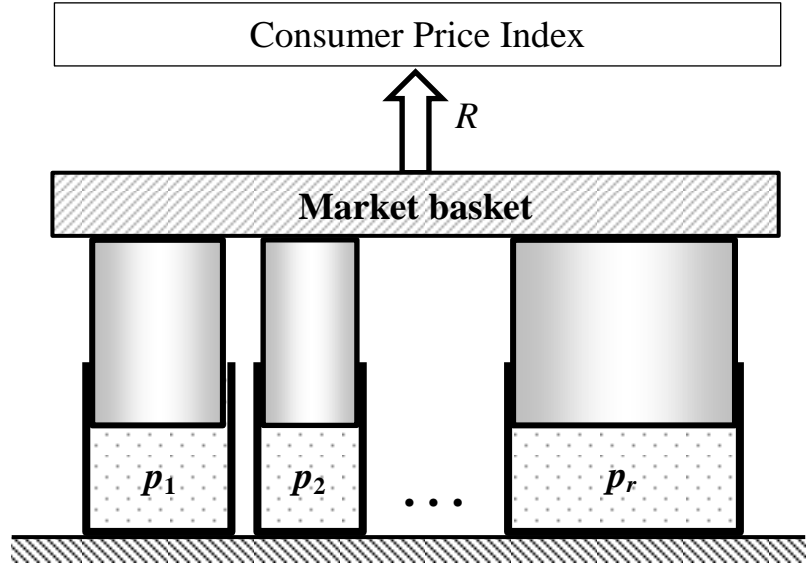

Fig. 1. Analogy between the market basket and a system of parallel pistons.

## 7. Thermodynamic conditions for market equilibrium stability

Consider a closed economic system ($N = const$) at a constant economic temperature ($T = const$) and at a constant pressure – the mean price of goods ($p = const$). Then inequality (64) takes the form

$$dG \leq 0 \qquad (107)$$

From (107), it follows that for fixed values of the parameters $T, p$ and $N$, the Gibbs free energy (63) of an economic system in a nonequilibrium state decreases monotonically and tends to the minimum value (65) corresponding to the equilibrium state of this system.

This means that the state of the economic system for the given parameters $T, p$ and $N$ is stable: the system always returns to an equilibrium state if it was removed from it by an external short-term influence.

Consider two states of the economic system corresponding to the same parameters $T, p$ and $N$: (i) the equilibrium state in which the energy of the system is $E$, the quantity of goods (volume of the system) is $V$, the entropy of the system is $S$ and the corresponding Gibbs free energy is $G$, and (ii) a non-equilibrium state in which the energy of the system is equal to $E_1$, the quantity of goods is equal to $V_1$, the entropy of the system is equal to $S_1$ and the corresponding Gibbs free energy is equal to $G_1$. Then, according to (107), $\Delta G = G_1 - G > 0$, or, taking into account (63),

$$E_1 - E + p(V_1 - V) - T(S_1 - S) > 0 \qquad (108)$$

Consider another equilibrium state of this economic system, in which it has the parameters $V_1, E_1$ and $S_1$. This state corresponds to other parameters $T_1, p_1$ and $N_1$. With the same parameters $T_1, p_1, N_1$ and with parameters $V, E, S$ the state of this economic system will be nonequilibrium. Then, by analogy with (108), one obtains

$$E - E_1 + p_1(V - V_1) - T_1(S - S_1) > 0 \qquad (109)$$

Folding inequalities (108) and (109), one obtains the general thermodynamic condition for the stability of equilibrium [20], as applied to the economic system

$$\Delta T \Delta S - \Delta p \Delta V > 0 \qquad (110)$$

where $\Delta T = T_1 - T$; $\Delta S = S_1 - S$; $\Delta p = p_1 - p$ and $\Delta V = V_1 - V$.

In particular, assuming $\Delta T = 0$, or $\Delta p = 0$, or $\Delta V = 0$, one obtains

$$\left(\frac{\Delta p}{\Delta V}\right)_T < 0, \ \left(\frac{\Delta p}{\Delta V}\right)_S < 0, \ \left(\frac{\Delta T}{\Delta S}\right)_p > 0, \ \left(\frac{\Delta T}{\Delta S}\right)_V > 0 \qquad (111)$$

Considering the infinitesimal variations of $\Delta V$ and $\Delta S$, from relations (111), one obtains the thermodynamic inequalities

$$\left(\frac{\partial p}{\partial V}\right)_T < 0, \ \left(\frac{\partial p}{\partial V}\right)_S < 0, \ \left(\frac{\partial T}{\partial S}\right)_p > 0, \ \left(\frac{\partial T}{\partial S}\right)_V > 0 \qquad (112)$$

which are the sufficient conditions for the stability of the equilibrium of the economic system.

The first two of conditions (111) and (112) have a simple economic meaning: an increase in the quantity of goods on the market at a constant economic temperature of the system or at a constant economic entropy (i.e., adiabatically, in the economic sense) leads to a decrease in the mean price of a good. These inequalities are a mathematical expression of the economic law of supply and demand [1]. Thus, the law of supply and demand is a natural and formal result of economic thermodynamics.

Taking into account the second law (4) and relations (25), one writes the stability conditions (111) in the form

$$\left(\frac{\Delta T}{\Delta S}\right)_p = T\left(\frac{\Delta T}{\delta Q}\right)_p = \frac{T}{C_p} > 0, \ \left(\frac{\Delta T}{\Delta S}\right)_V = T\left(\frac{\Delta T}{\delta Q}\right)_V = \frac{T}{C_V} > 0 \qquad (113)$$

Taking into account that the economic temperature is positive by definition, from relations (113) one obtains

$$C_V > 0, \ C_p > 0 \qquad (114)$$

Thus, for the stability of the economic system, its heat capacity at constant volume and at constant pressure (goods price) must be positive.

In a similar way, considering the Grand potential (67), one obtains another general thermodynamic condition for the stability of equilibrium in relation to the economic system:

$$\Delta T \Delta S + \Delta \mu \Delta N > 0 \qquad (115)$$

from which the conditions follow

$$\left(\frac{\Delta \mu}{\Delta N}\right)_T > 0, \ \left(\frac{\Delta \mu}{\Delta N}\right)_S > 0, \ \left(\frac{\Delta T}{\Delta S}\right)_N > 0, \ \left(\frac{\Delta T}{\Delta S}\right)_\mu > 0 \qquad (116)$$

and

$$\left(\frac{\partial \mu}{\partial N}\right)_T > 0, \ \left(\frac{\partial \mu}{\partial N}\right)_S > 0, \ \left(\frac{\partial T}{\partial S}\right)_N > 0, \ \left(\frac{\partial T}{\partial S}\right)_\mu > 0 \qquad (117)$$

where $\Delta \mu = \mu_1 - \mu$ and $\Delta N = N_1 - N$.

## 8. Le Chatelier's principle in economics

It is obvious that already by virtue of the definition of stability, a stable system "resists" any changes caused by both internal and external influences.

Let any parameter $x = (p, V, N, E, S^0)$ of the system changed by a small value $x_0$, under the action of external influences,. This will lead to a change in the state of the system, and will cause internal changes in it. As a result of these internal processes, the parameter $x$ will change additionally by the value $\delta x$. Because the change $\delta x$ is caused by a change $\delta x_0$, one can write

$$\delta x = g \delta x_0 \qquad (118)$$

where the factor $g$ is the sensitivity of the system to a change in the parameter $x$ in the process under consideration.

The total change in the parameter $x$ in this process is

$$\Delta x = \delta x_0 + \delta x = (1 + g)\delta x_0 \qquad (119)$$

Obviously, if the system is stable, then the condition

$$|\Delta x| < |\delta x_0| \qquad (120)$$

since otherwise, the total change $\Delta x$, in turn, will cause an even greater deviation $\delta x$, and self-amplification of the initial perturbation $\delta x_0$ will occur, which is incompatible with the concept of system stability.

It follows from (118) - (120) that the condition for the stability of the equilibrium state of the system is

$$-2 \leq g \leq 0 \qquad (121)$$

Obviously, at $-1 < g < 0$, the system returns from an excited (nonequilibrium) state to an equilibrium state monotonically (without oscillations, in the form of relaxation). If $-2 < g <$ $-1$, then the process of returning the system from an excited (nonequilibrium) state to an equilibrium state occurs in the form of damped oscillations.

Thus, a change $\delta x$ in a stable system caused by internal processes induced by an external action is always directed against a change $\delta x_0$ caused by this external action.

This allows formulating the Le Chatelier's principle.

*If the system is in a stable state, then any external action causes internal changes in it that counteract this action; that is, internal changes occurring in the system tend to reduce the result of external action.*

Note that Le Chatelier's principle is valid only for systems in an equilibrium (i.e., asymptotically stable) state. If the system is unstable, then the perturbations that always take place will bring the system out of the state it is in, and the system will no longer return to this state, but will "search" for a new stable state, if it exists.

In chemistry, the Le Chatelier principle is one of the guiding principles that allows one to quickly, without a detailed analysis of the process, establish its direction, as well as understand what changes in the system will lead to certain effects on it.

The Le Chatelier principle plays a similar role in economics (and, in general, in social systems): in many cases, based only on this principle, without a detailed economic analysis, it is possible to predict what changes will occur in the economic system under the influence of one or another possible change or impact.

In economics, Le Chatelier's principle was introduced in [25].

Thus, factor-demand and commodity-supply elasticities are hypothesized to be lower in the short run than in the long run because of the fixed-cost constraint in the short run. [25-27].

It was shown [26] that Le Chatelier's principle is a corollary of the envelope theorem [28].

Consider examples of the application of the Le Chatelier's principle to economic systems.

1. A change in the economic temperature of a stable system due to some impact will cause such structural changes in the system that lead to a change in the economic temperature in the opposite direction. According to (19), the economic temperature of the system is equal to the amount of money available in the system per one element of the system. Taking this definition into account, it follows from the Le Chatelier's principle that when the economic temperature of the system rises, people begin to invest in real estate, stocks, investments, etc., i.e. strive to perform those actions that lead to a decrease in the amount of free money in the system, and, therefore, to a decrease in the economic temperature. On the contrary, when the economic temperature of the system decreases, people tend to sell real estate, shares, etc., i.e. their actions are aimed at increasing the amount of free money in the system, and therefore, to increase in the economic temperature of the system.

2. A decrease in the volume of the system (quantity of goods) will lead, according to conditions (111) and (112), to an increase in pressure (price of goods) in the system. In this case, the economic temperature of the system will change in such a direction as to increase the volume of the system again.

Indeed, let in some elementary process there was a change in the volume of the system $\delta V_0$, which led to a change in the economic temperature

$$\delta T = \left(\frac{\partial T}{\partial V}\right) \delta V_0 \qquad (122)$$

where the derivative is taken under conditions appropriate to the process under consideration. At the same time, a change in the economic temperature of the system at constant pressure (price of goods) leads to a change in the volume of the system due to its "thermal" expansion:

$$\delta V = \left(\frac{\partial V}{\partial T}\right)_p \delta T \qquad (123)$$

Taking into account (122), one writes (123) in the form

$$\delta V = \left(\frac{\partial V}{\partial T}\right)_p \left(\frac{\partial T}{\partial V}\right) \delta V_0 \qquad (124)$$

Comparing (124) with (118), in this case one obtains

$$g = \left(\frac{\partial V}{\partial T}\right)_p \left(\frac{\partial T}{\partial V}\right) \qquad (125)$$

For a stable system, parameter (125) must satisfy condition (121), i.e. $g < 0$.

Therefore, systems (markets) that are compressed when "heated" ($\left(\frac{\partial V}{\partial T}\right)_p < 0$) will "cool" when compressed, i.e. as the market volume decreases, its economic temperature decreases ($\left(\frac{\partial T}{\partial V}\right) > 0$), while systems that expand when "heated" ($\left(\frac{\partial V}{\partial T}\right)_p > 0$) will "heat up" when compressed, i.e. as the market volume decreases, its economic temperature increases ($\left(\frac{\partial T}{\partial V}\right) < 0$).

3. Suppose the government decides to increase the personal income tax in the hope of replenishing the budget. If it does this, then people will have less money for their own needs. As a result, people will abandon some of the purchases (expenses) they used to make; the respective firms will receive less income and, therefore, will pay less tax to the state than before the income tax increase. Moreover, the increase in tax will cause the people and businesses to look for ways (legal and sometimes illegal) to evade taxes. All this will lead to the fact that the result from the tax increase will be less than expected. If the result of the tax increase turns out to be negative, i.e. the total income of money to the budget after the tax increase will decrease, the government will be forced to reduce the tax.

4. Suppose that all sellers of a certain good in the market decide to increase its price in order to increase their income. If they do this, then some buyers will either stop buying this good (for example, completely abandon it), or buy it in smaller quantities. This will lead to a decrease in consumer demand and market turnover, as a result of which the total income of sellers of this good will be less than it could have been if the market turnover had not changed. If the change in the price of a good was abrupt, then the drop in demand for it will also be abrupt. This will force sellers to slightly lower the price of the good again (which may nevertheless be higher than the original price) in order to increase demand.

## 9. Interaction of two markets

By definition, two interacting systems are in equilibrium if their interaction does not lead to a change in their states.

Consider two interacting markets in different states, i.e. it is believed that economic temperatures, prices for the same good, quantity of good (the volume of markets) and number of elements in these markets are different. We assume that these markets can interact (i.e. exchange goods, money and elements) only with each other. In this case, we can consider a combined system consisting of these two markets (two subsystems), which is isolated, i.e. for it $E = E_1 + E_2 = const, N = N_1 + N_2 = const, V = V_1 + V_2 = const$. Then

$$dE_1 + dE_2 = 0, dV_1 + dV_2 = 0, dN_1 + dN_2 = 0 \qquad (126)$$

Taking into account the first law (1), the change in the entropy of the combined system is described by the equation [19]

$$dS = \frac{1}{T_1}dE_1 + \frac{1}{T_2}dE_2 + \frac{p_1}{T_1}dV_1 + \frac{p_2}{T_2}dV_2 - \frac{\mu_1}{T_1}dN_1 - \frac{\mu_2}{T_2}dN_2 \qquad (127)$$

Since the systems under consideration have different intensive parameters, they are not in equilibrium, and the combined system is also not in equilibrium. Thus, for a combined adiabatically isolated system, the inequality $dS \geq 0$ takes place, which follows from the second law (2) at $\delta Q = 0$.

Taking into account (126) and (127), one obtains

$$\left(\frac{1}{T_1} - \frac{1}{T_2}\right)dE_1 + \left(\frac{p_1}{T_1} - \frac{p_2}{T_2}\right)dV_1 - \left(\frac{\mu_1}{T_1} - \frac{\mu_2}{T_2}\right)dN_1 \geq 0 \qquad (128)$$

Since the parameters $E, V, N$ can change independently, we come to the conclusion that in a state of equilibrium (the equal sign in (128))

$$T_1 = T_2, p_1 = p_2, \mu_1 = \mu_2 \qquad (129)$$

Conditions (129) are equilibrium conditions for two interacting markets.

Thus, we have rigorously proved that two interacting markets are in equilibrium if they have the same intensive parameters: economic temperature, pressure (commodity price) and financial potential [19].

If at least one of the conditions (129) is not met, then the interacting markets are in a nonequilibrium state, and the exchange of money, goods and elements (market actors) will occur between them until an equilibrium is established, corresponding to condition (129).

In the general case, changes in the energy, volume and number of market elements can occur interconnected. For example, the difference in pressures (prices for the same goods) between systems can lead not only to a change in the volumes of these systems (the amount of goods in them), but also to a change in the number of elements due to the migration of elements from one system to another, etc. This issue will be addressed in future papers in this series.

In the simplest case, considering the change in only one of the extensive parameters of the market (energy, volume or number of elements), using inequality (128), it is possible to establish the direction of the economic process.

For example, consider the process of heat transfer between markets when

$$dV_1 = dV_2 = 0, dN_1 = dN_2 = 0 \tag{130}$$

This means that markets cannot sell goods to each other, and elements of markets cannot move from one system to another. However, these markets have the ability to directly exchange money (heat), i.e. participants in one market can transfer money to participants in another market (donate, invest, etc.).

This case was considered in [19], where it was shown that heat (in the economic sense) always spontaneously flows from a system with a higher economic temperature to a system with a lower economic temperature.

In this case, the energy of the markets can only change due to heat exchange (in the above sense) between them:

$$dE_2 = -dE_1 \neq 0 \tag{131}$$

and inequality (128) takes the form

$$\left(\frac{1}{T_1} - \frac{1}{T_2}\right) dE_1 > 0 \tag{132}$$

Hence it follows that for $T_1 < T_2$ there should be

$$dE_1 > 0 \tag{133}$$

i.e. under conditions (130), the energy of a system with a high economic temperature decreases, while the energy of a system with a lower economic temperature increases. Equilibrium occurs when $T_1 = T_2$. In order for the systems to reach a state of equilibrium, and this equilibrium was stable, the temperature of a system with a higher temperature must increase, while the temperature of a system with a lower temperature must decrease. Thus, we come to the conclusion that for the stability of the economic system (market), its energy must increase with increasing temperature. Taking into account (36), this means that condition (114), which was obtained above from other considerations, must be satisfied for a stable economic system.

The simple example of heat transfer between economic systems is the process when people living in richer countries (cities) with high economic temperatures transfer money en masse to their relatives living in poorer countries (cities) with low economic temperatures. For the same reason, other things being equal, investments are always directed from richer countries (with high economic temperatures) to poorer countries (with low economic temperatures) [9].

Consider in a similar way two interacting systems for which

$$T_1 = T_2, p_1 \neq p_2, \mu_1 = \mu_2 \tag{134}$$

and which can sell goods to each other ($dV_1 = -dV_2 \neq 0$).

In this case, taking into account that $T > 0$, inequality (128) takes the form

$$(p_1 - p_2)dV_1 \geq 0 \tag{135}$$

It follows that the volume of the market with a higher price of the goods (more pressure) increases, while the volume of the market with a lower price of the same good (less pressure) decreases; i.e. a market with higher pressure always expands at the expense of a market with less pressure.

As an example, consider two markets for the same good. Their volumes are $V_1$ and $V_2$. The prices for this good are $p_1$ and $p_2$, respectively. All other things being equal, these markets are in equilibrium if $p_1 = p_2$. Let, for some reason, become $p_1 \neq p_2$, all other things being equal. For definiteness, we take $p_1 > p_2$. That is, for some reason, the same good in market 1 costs more than in market 2. Obviously, in this case, market 1 will start buying goods in market 2. As a result, the volume of market 1 will increase, while the volume of market 2 will decrease at $V = V_1 + V_2 = const$. Thus, in fact, the expansion of the system with higher pressure (market 1) occurs due to the compression of the system with lower pressure (market 2), which is proved by relation (135). According to the thermodynamic conditions of the market equilibrium stability (136) and (137), an increase in the volume of goods in market 1 leads to a decrease in the price $p_1$, while a decrease in the volume of goods in market 2 leads to an increase in the price $p_2$. This process will continue until the prices for this good in markets 1 and 2 equalize. This result is remarkable. From the point of view of ordinary economic theory [1], it is trivial, but there it was obtained not as a result of a rigorous mathematical proof, but, in fact, on the basis of general reasoning based on considerations of a psychological nature. In the thermodynamic theory of markets, we have obtained this result as a strict formal consequence of the theory. This indicates that, apparently, many results of economic theory, which, in fact, are based on the analysis of human psychology, in economic thermodynamics can receive a rigorous mathematical (formal) justification.

A direct thermodynamic analogy of the system under consideration is a cylindrical vessel filled with gas and divided into two parts by a movable piston: the piston moves until the pressure on both its sides equalizes.

Note that the two systems can exchange goods (buy and sell goods to each other) both directly and through the third system. For example, suppose there are two banks that buy and sell Euros for US dollars. In this case, US dollars play the role of money (energy), while Euros plays the role of goods (although the opposite point of view is also possible: Euros plays the role of money while US dollars play the role of goods). Suppose banks cannot directly buy or sell Euros to each other, but can buy Euros from the individuals or sell Euros to the individuals. Let bank 1 buy Euros at $p_1$, and bank 2 sells Euros at $p_2$. Assume that for some reason, $p_1 > p_2$. In this case, individuals will begin to massively buy Euros in bank 2, immediately sell them in bank 1, and again buy Euros in bank 2 with the dollars earned. The process will continue until the Euro-to-dollar rates in these banks equalize. In this case, the individuals play the role of a third system ("piston") through which banks interact.

Consider in a similar way two interacting systems for which

$$T_1 = T_2, p_1 = p_2, \mu_1 \neq \mu_2 \qquad (136)$$

and which can exchange elements ($dN_1 = -dN_2 \neq 0$).

In this case, taking into account that $T > 0$, inequality (128) takes the form

$$(\mu_2 - \mu_1)dN_1 \geq 0 \qquad (137)$$

It follows that the number of elements in the system (on the market) with a lower financial potential $\mu$ increases due to the migration of elements from the system (market) with a higher financial potential $\mu$. This process is similar to diffusion in physical systems.

According to the first two market equilibrium stability conditions (116) and (117), an increase in the number of elements in the system, all other things being equal, leads to an increase in the financial potential of the system. Thus, the migration of elements from a system with a lower financial potential to a system with a higher financial potential will continue until their financial potentials equalize. In this case, migration (diffusion) equilibrium will be established between the systems. This means that elements can still move from one system to another, however, on average, how many elements passed from one system to another, the same number of elements moved in the opposite direction.

In a more general case, when $T_1 \neq T_2$, from the general condition (128) instead of inequality (137) one obtains

$$\left(\frac{\mu_2}{T_2} - \frac{\mu_1}{T_1}\right) dN_1 \geq 0 \tag{138}$$

As an example, suppose interacting markets 1 and 2 are primitive (ideal). In this case, their financial potentials are described by relations (83) and (84). Then inequality (138) takes the form

$$(\ln x_2 - \nu \ln T_2 - \ln x_1 + \nu \ln T_1) dN_1 \geq 0 \tag{139}$$

Here, for simplicity, it is assumed that the number of financial degrees of freedom in both markets is the same: $\nu_1 = \nu_2 = \nu$, and also $B_1 = B_2$.

From condition (139) it follows that at $T_1 = T_2$ the elements will move (migrate, diffuse) from the market, where their "concentration" $x$ is greater, to the market, where their "concentration" is less, which corresponds to the conditions of diffusion in physical systems.

If the "concentrations" of elements in both markets are the same ($x_1 = x_2$), but their economic temperatures are different, then, according to (139), migration (diffusion) of elements occurs from a market with a lower economic temperature to a market with a higher economic temperature. This process can be called economic thermodiffusion (migration of elements under the influence of the difference in economic temperatures).

It is easy to verify that, as a result of economic thermodiffusion, the systems under consideration tend to a state of equilibrium.

Indeed, due to thermodiffusion, the number of elements in a system with a higher temperature increases, while in a system with a lower temperature, it decreases. If the amount of money in these systems does not change, then an increase in the number of elements leads to a decrease in the economic temperature in the "overheated" system and to an increase in the economic temperature in the "colder" system. If, at the same time, heat exchange takes place between the systems, then the process of equalizing the economic temperatures of the interacting systems will proceed even faster. All this leads to the disappearance of the cause of thermodiffusion.

Obviously, the described economic thermodiffusion is fully consistent with our ideas that, other things being equal, migration is always directed from poorer economic systems (with a low economic temperature) to richer systems (with a higher economic temperature).

Note that, in contrast to thermal diffusion in physical systems, which, most often is directed from a system with a higher temperature to a system with a lower temperature, economic thermodiffusion (at least between primitive (ideal) markets) is directed from a system with a lower economic temperature to a system with a higher economic temperature.

## 10. Concluding remarks

Thus, in [19] and in the present work, we have shown that economic thermodynamics (thermodynamic method for describing economic systems and processes) can be constructed as a phenomenological theory, by analogy with how it is done in physics.

Table 1. Thermodynamic analogies in economics

| Parameter, process | Thermodynamic system | Economic system |
|---|---|---|
| $E=K+U$ | Internal energy | Money, financial instruments, property (movable and immovable), etc. |
| $K$ | Kinetic energy of atoms and molecules | Money and financial instruments of market actors playing the role of money |
| $U$ | Potential energy of interaction of particles inside the system and with external objects (fields) | Material resources and property (movable and immovable) of market actors. |
| $V$ | Volume | Quantity of goods |
| $p$ | Pressure | Mean price of a unit of goods |
| $T$ | Thermodynamic temperature | Economic temperature – the amount of money in the system per one element of the system. |
| $N$ | The number of system elements – atoms and molecules | Number of system elements – market actors |
| $\mu$ | Chemical potential | Financial potential |
| $\delta Q$ | Thermal energy, heat | The amount of money that elements transfer to each other or external systems without buying and selling goods |
| $\delta W = pdV$ | Mechanical work | Changing the amount of money in the system due to the purchase and sale of goods. |
| Dissociation-recombination | Dissociation-recombination of atoms and molecules, chemical reactions | Formation and disintegration of families, firms, companies, etc. |
| Thermal conductivity, heat transfer | Transfer of heat energy between systems due to temperature differences in systems | Transfer of money from one system to another, due to the different economic temperatures in the systems |
| Diffusion | Diffusion of atoms and molecules due to the difference in their concentration in systems | Migration of elements from one system to another due to the difference in concentrations (in a generalized sense) of elements in systems |
| Thermodiffusion, thermophoresis | Diffusion of atoms and molecules associated with temperature differences between systems | Migration of elements between systems due to the difference in economic temperatures between systems. |
| … | … | … |

We see that there is a deep analogy between the parameters of thermodynamic and economic systems (markets). In particular, each thermodynamic parameter can be associated with a certain economic parameter or indicator (see Table 1).

Obviously, this table of analogies is incomplete. There are many analogies between thermodynamic and economic systems that have yet to be established. I left the last line blank, in the belief that the list of analogies will be significantly expanded in the future.

The results obtained in [19] and in this work give hope that many economic laws, which in economic theory are a generalization of the results of observations, or are based on the analysis of the psychology of the behavior of market actors, within the framework of economic thermodynamics can be obtained as the natural and formal results of the theory. So, for example, we see that slow (quasi-static) inflation is naturally described by economic thermodynamics.

In [19] and in this work, we considered only two types of interaction of economic systems (or their elements): material (exchange of elements) and energy (exchange of money). However, in economic systems, in contrast to physical thermodynamic systems, there is another type of interaction that significantly affects the processes occurring in these systems: information interaction or information exchange. It seems that this type of interaction should also be included in economic thermodynamics (for example, by generalizing the first and second laws). We hope to discuss this issue in future papers in this series.

**CRediT authorship contribution statement**

**S.A. Rashkovskiy:** Conceptualization, Methodology, Writing - original draft, Investigation, Writing - review & editing.

**Declaration of competing interest**

The authors declare that they have no known competing financial interests or personal relationships that could have appeared to influence the work reported in this paper.

**Acknowledgments**